\documentclass{elsart}

\usepackage{epsfig}
\usepackage{amssymb}

\begin{document}

\begin{frontmatter}

\title{Wormhole inspired by non-commutative geometry}

\author[label1]{Farook Rahaman}\ead{rahaman@iucaa.ernet.in},
\author[label2]{Sreya Karmakar}\ead{sreya.karmakar@gmail.com},
\author[label3]{Indrani Karar}\ead{indrani.karar08@gmail.com},
\author[label4]{Saibal Ray}\ead{saibal@iucaa.ernet.in}

\address[label1]{Department of Mathematics, Jadavpur University,
Kolkata 700032, West Bengal, India}
\address[label2]{Department of Physics, Calcutta Institute of
Engineering and Management, Kolkata 700040, West Bengal, India}
\address[label3]{Department of Mathematics, Saroj Mohan
Institute of Technology, Guptipara, West Bengal, India}
\address[label4]{Department of Physics, Government College of Engineering \& Ceramic
Technology, Kolkata 700010, West Bengal, India}

\begin{abstract}
In the present letter we search for a new wormhole solution inspired
by noncommutative geometry with the additional condition of
allowing conformal Killing vectors (CKV). A special aspect of
noncommutative geometry is that it replaces point-like structures
of gravitational sources with smeared objects under Gaussian
distribution. However, the purpose of this paper is to obtain
wormhole solutions with noncommutative geometry as a background
where we consider a point-like structure of gravitational object
without smearing effect. It is found through this investigation
that wormhole solutions exist in this Lorentzian distribution with
viable physical properties.
\end{abstract}


\end{frontmatter}

\section{INTRODUCTION}

A {\it wormhole}, which is similar to a tunnel with two ends each
in separate points in spacetime or two connecting black holes, was
conjectured first by Weyl \cite{Coleman1985} and later on by
\cite{Wheeler1957}. In a more concrete physical definition it is
essentially some kind of hypothetical topological feature of
spacetime which may acts as {\it shortcut} through spacetime
topology.

It is argued by Morris et al. \cite{Morris1988a,Morris1988b} and
others \cite{Ellis1973,Bronnikov1973,Clement1984} that in
principle a wormhole would allow travel in time as well as in
space and can be shown explicitly how to convert a wormhole
traversing space into one traversing time. However, there are
other types of wormholes available in the literature where the
traversing path does not pass through a region of exotic matter
\cite{Visser1989,Visser1995}. Following the work of Visser
\cite{Visser1989}, a new type of thin-shell wormhole, which was
constructed by applying the cut-and-paste technique to two copies
of a charged black hole \cite{Usmani2010}, is of special mention
in this regard.

Thus a traversable wormhole, tunnel-like structure connecting
different regions of our Universe or of different universes
altogether, has been an issue of special investigation under
Einstein's general theory of relativity \cite{Rahaman2013a}. It is
argued by Rahaman et al. \cite{Rahaman2014a} that although just as
good a prediction of Einstein's theory as black holes, wormholes
have so far eluded detection. As one of the peculiar features a
wormhole requires the violation of the null energy condition (NEC)
\cite{Morris1988b}. One can note that phantom dark energy also
violates the NEC and hence could have deep connection to in
formation of wormholes \cite{Lobo2005,Kuh2009}.

It is believed that some perspective of quantum gravity can be
explored mathematically in a better way with the help of
non-commutative geometry. This is based on the non-commutativity
of the coordinates encoded in the commutator, $[x_{\mu},x_{\nu}] =
\theta_{\mu\nu}$, where $\theta_{\mu\nu}$ is an anti-symmetric and
real second-ordered matrix which determines the fundamental cell
discretization of spacetime
\cite{Doplicher1994,Kase2003,Smailagic2004,Nicolini2006,Nicolini2009}.
We also invoke the inheritance symmetry of the spacetime under
conformal Killing vectors (CKV). Basically CKVs are motions along
which the metric tensor of a spacetime remains invariant up to a
certain scale factor. In a given spacetime manifold $M$, one can
define a globally smooth conformal vector field $\xi$, such that
for the metric $g_{ab}$ it can be written as
\begin{equation}
 \xi_{a;b} = \psi g_{ab} + F_{ab},
\end{equation}
where $\psi: M \rightarrow R$ is the smooth conformal function of
$\xi$ and $F_{ab}$ is the conformal bivector of $\xi$. This is
equivalent to the following form:
\begin{equation}
L_{\xi} g_{ik} =\xi_{i;k}+ \xi_{k;i} = \psi g_{ik},\label{ckv}
\end{equation}
where $L$ signifies the Lie derivatives along the CKV
$\xi^{\alpha}$.

In favor of the prescription of this mathematical technique CKV we
find out the following features: (1) it provides a deeper insight
into the spacetime geometry and facilitates the generation of
exact solutions to the Einstein field equations in a more
comprehensive forms, (2) the study of this particular symmetry in
spacetime is physically very important as it plays a crucial role
of discovering conservation laws and to devise spacetime
classification schemes, and (3) because of the highly
non-linearity of the Einstein field equations one can reduce
easily the partial differential equations to ordinary differential
equations by using CKV. Interested readers may look at the recent
works on CKV technique available in the literature
\cite{Rahaman2013b,Rahaman2014b,Rahaman2014c}.

In this letter therefore we search for some new solutions of
wormhole admitting conformal motion of Killing Vectors. It is a
formal practice to consider the inheritance symmetry to establish
a natural relationship between spacetime geometry and
matter-energy distribution for a astrophysical system. Thus our
main goal here is to examine the solutions of Einstein field
equations by admitting CKV under non-commutative geometry. The
scheme of the investigation is as follows: in the Sec. 2 we
provide the mathematical formalism and Einstein's field equations
under the framework of this technique. A specific matter-energy
density profile has been employed in Sec. 3 to obtain various
physical features of the wormhole under consideration by
addressing the issues like the conservation equation, stability of
the system, active gravitational mass and gravitational energy.
Sec. 4 is devoted for some concluding remarks.

\section{CONFORMAL KILLING VECTOR AND BASIC EQUATIONS}

We take the static spherically symmetric metric in the following
form
\begin{equation}
ds^2=e^{\nu(r)} dt^2 - e^{\lambda(r)} dr^2 -
r^2(d\theta^2+sin^2\theta d\phi^2),
\end{equation}
where $r$ is the radial coordinate. Here $\nu$ and $\lambda$ are
the metric potentials which have functional dependence on $r$
only.

Thus, the only survived Einstein's field equations in their
explicit forms (rendering $G = c = 1$) are
\begin{equation}
e^{-\lambda}\left[ \frac{\lambda'}{r}-\frac{1}{r^2}\right]
+\frac{1}{r^2}= 8\pi \rho,
\end{equation}

\begin{equation}
e^{-\lambda}\left[ \frac{1}{r^2}+\frac{\nu'}{r}\right]
-\frac{1}{r^2}= 8\pi p_r,
\end{equation}

\begin{equation}
\frac{1}{2} e^{-\lambda}\left[
\frac{1}{2}(\nu')^2+\nu''-\frac{1}{2}\lambda' \nu' +
\frac{1}{r}(\nu'-\lambda')\right] = 8\pi p_t,
\end{equation}
where $\rho$, $p_r$ and $p_t$ are matter-energy density, radial
pressure and transverse pressure respectively for the fluid
distribution. Here $\prime$ over $\nu$ and $\lambda$ denotes
partial derivative w.r.t. radial coordinate $r$.

The conformal Killing equations, as mentioned in Eqs. (\ref{ckv}),
then yield as follows:

$  \xi^1 \nu^\prime =\psi $,

$ \xi^4  = C_1 = constant$,

$ \xi^1  = \frac{\psi r}{2}$,

$ \xi^1 \lambda ^\prime + 2 \xi^1 _{,1} =\psi $,

\noindent where $\xi^{\alpha}$ are the conformal $4$-vectors and
$\psi$ is the conformal function as mentioned earlier.

This set of equations, in a straight forward way, imply the
following simple forms:

\begin{equation}
e^\nu  =C_2^2 r^2,
\end{equation}

\begin{equation}
e^{\lambda}  = \frac{{C_3}^2}{{\psi}^2},
\end{equation}

\begin{equation}
\xi^i = C_1 \delta_4^i + \left(\frac{\psi r}{2}\right)\delta_1^i,
\end{equation}
where $C_2$ and $C_3$ are integration constants. Here the non-zero
components of the conformal Killing vector $\xi^a$ are $\xi^0$ and
$\xi^1$.

Now using solutions (7) and (8), the equations (3)-(5) take the
following form as

\begin{equation}
\frac{1}{r^2}\left[1 - \frac{\psi^2}{C_3^2}
\right]-\frac{2\psi\psi^\prime}{C_3^2 r}= 8\pi \rho,
\end{equation}

\begin{equation}\frac{1}{r^2}\left[1 - \frac{3\psi^2}{C_3^2}
\right]= - 8\pi p_r,
\end{equation}

\begin{equation}\left[\frac{\psi^2}{C_3^2r^2}
\right]+\frac{2\psi\psi^\prime}{C_3^2r} =8\pi p_t.
\end{equation}

These are the equations forming master set which has all the
information of the fluid distribution under the framework of
Einstein's general theory of relativity with the associated
non-commutative geometry and conformal Killing vectors.

\section{THE MATTER-ENERGY DENSITY PROFILE AND PHYSICAL FEATURES OF THE WORMHOLE}

As stated by Rahaman et al. \cite{Rahaman2013a}, the necessary
ingredients that supply fuel to construct wormholes remain an
elusive goal for theoretical physicists and there are several
proposals that have been put forward by different authors
\cite{Kuh1999,Sushkov2005,Lobo2006,Rahaman2008,Rahaman2009,Kuh2010}.
However, in our present work we consider cosmic fluid as source
and thus have provided a new class of wormhole solutions. Keeping
the essential aspects of the noncommutativity approach which are
specifically sensitive to the Gaussian nature of the smearing as
employed by Nicolini et al. \cite{Nicolini2006}, we rather get
inspired by the work of Mehdipour \cite{Mehdipour2012} to search
for a new fluid model admitting conformal motion. Therefore, we
assume a Lorentzian distribution of particle-like gravitational
source and hence the energy density profile as given in Ref.
\cite{Mehdipour2012} as follows:
\begin{equation}
\rho(r) = \frac{M \sqrt{\phi}}{\pi^2(r^2 +\phi)^2},
\end{equation}
where $\phi$ is the noncommutativity parameter and $M$ is the
smeared mass distribution.

Now, solving equation (10) we get
\begin{equation}
\psi^2 = C_3^2-\left(\frac{4 C_3^2 M}{\pi
r}\right)\left[tan^{-1}\left(\frac{r}{\sqrt{\phi}}\right)-\frac{r\sqrt{\phi}}{r^2+\phi}\right]
+ \frac{D_1}{r},
\end{equation}
where $D_1$ is an integration constant and can be taken as zero.

The parameters, like the radial pressure, tangential pressure and
metric potentials, are found as
\begin{equation}
 p_r  = \frac{1}{8 \pi} \left[ \frac{2}{r^2} -   \frac{12M  }{\pi r^3}
 \left\{
tan^{-1}\left(\frac{r}{\sqrt{\phi}}\right)-\frac{r\sqrt{\phi}}{r^2
+\phi}\right\}\right],
\end{equation}
\begin{equation}
 p_t  = \frac{1}{8 \pi} \left[ \frac{1}{r^2} - 8 \pi \left(\frac{M \sqrt{\phi}}{\pi^2(r^2 +\phi)^2}
\right)\right],
\end{equation}

\begin{equation}
e^\nu = C_2^2 r^2,
\end{equation}

\begin{equation}
e^\lambda = \frac{1}{\left[1-\left(\frac{4M}{\pi
r}\right)\left(tan^{-1}\left(\frac{r}{\sqrt{\phi}}\right)-\frac{r\sqrt{\phi}}{r^2
+\phi}\right)\right]}.
\end{equation}

Let us now write down the metric potential conveniently in terms
of the shape function $b(r)$ as follows:
\begin {equation}
e^\lambda = \frac{1}{1-\frac{b(r)}{r}},
\end{equation}
where $b(r)$ is given by
\begin{equation}
b(r)= \left(\frac{4M}{\pi
}\right)\left[tan^{-1}\left(\frac{r}{\sqrt{\phi}}\right) - \frac{r
\sqrt{\phi}}{r^2+\phi}\right].
\end{equation}

\begin{figure}
      \centering   \includegraphics[scale=.50]{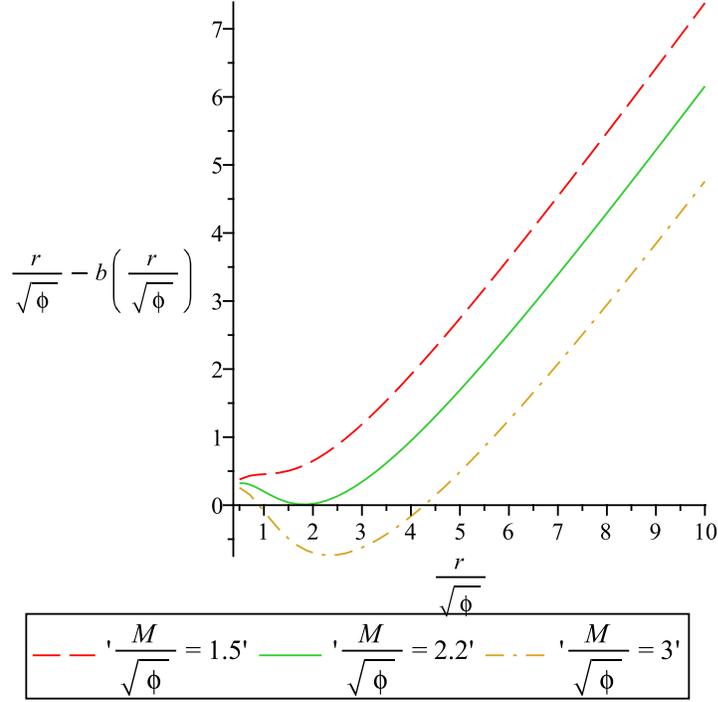}
        \caption{The throat of the wormhole is located at $\frac{r}{\sqrt{\phi}} =
\frac{r_0}{\sqrt{\phi}}$ (maximum root), where
$b\left(\frac{r}{\sqrt{\phi}}\right) -  \frac{r}{\sqrt{\phi}}$
cuts the   $\frac{r}{\sqrt{\phi}}$-axis. For
$\frac{M}{\sqrt{\phi}} < 2.2$, there exists no root, and therefore
no throats. For $\frac{M}{\sqrt{\phi}} > 2.2$, however we have two
roots: (i) for $\frac{M}{\sqrt{\phi}} = 3$, the location of the
external root i.e. throat of the wormhole is
$\frac{r_0}{\sqrt{\phi}}=4.275$, and (ii) for
$\frac{M}{\sqrt{\phi}} = 2.2$, we have one and only one solution
and this corresponds to the situation when two roots coincide and
it can be interpreted as an "extreme " situation}
   \label{fig:shape3}
\end{figure}

\begin{figure*}[thbp]
\begin{tabular}{rl}
\includegraphics[width=4.9cm]{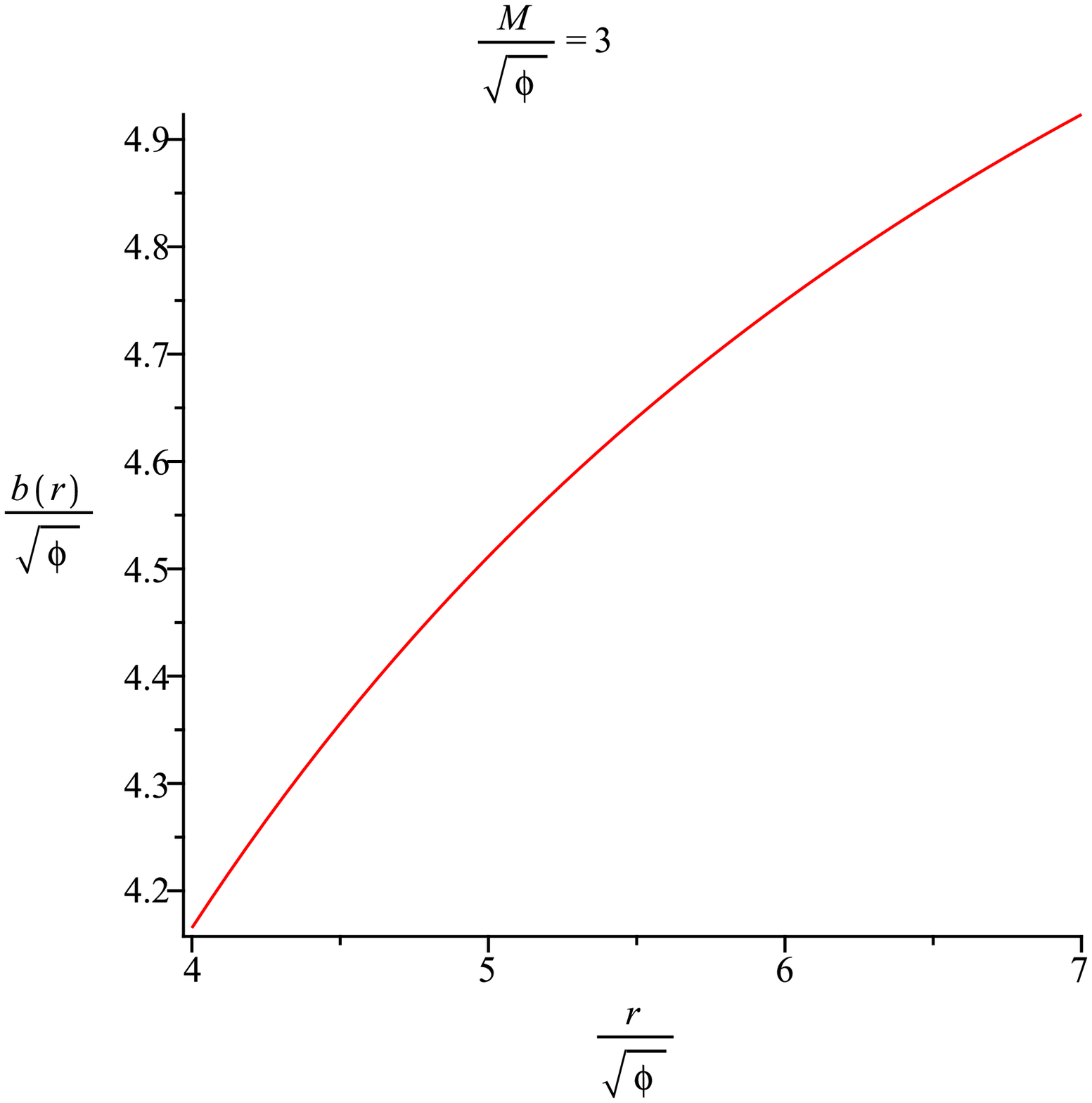}&
\includegraphics[width=4.9cm]{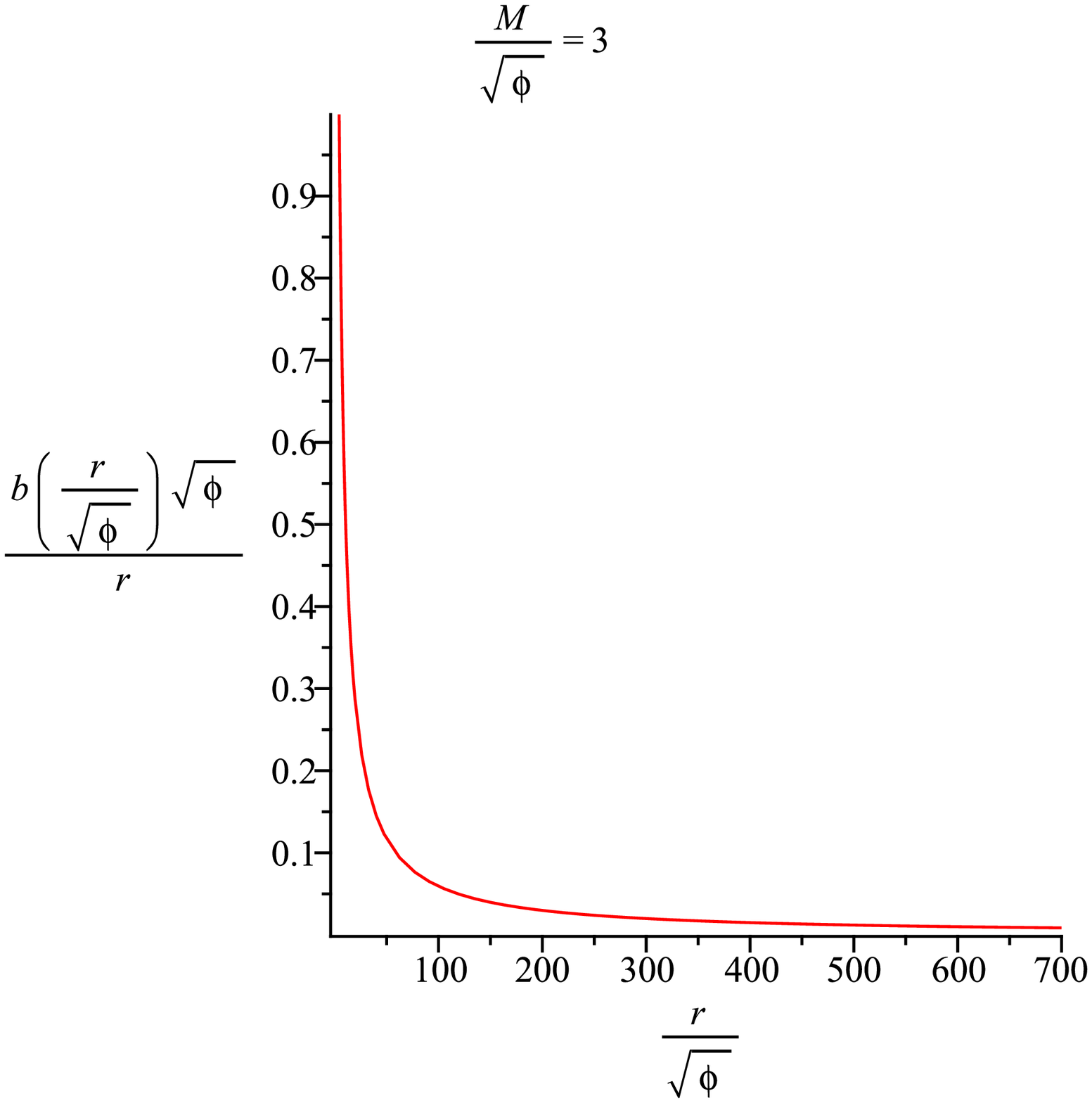}
\includegraphics[width=4.9cm]{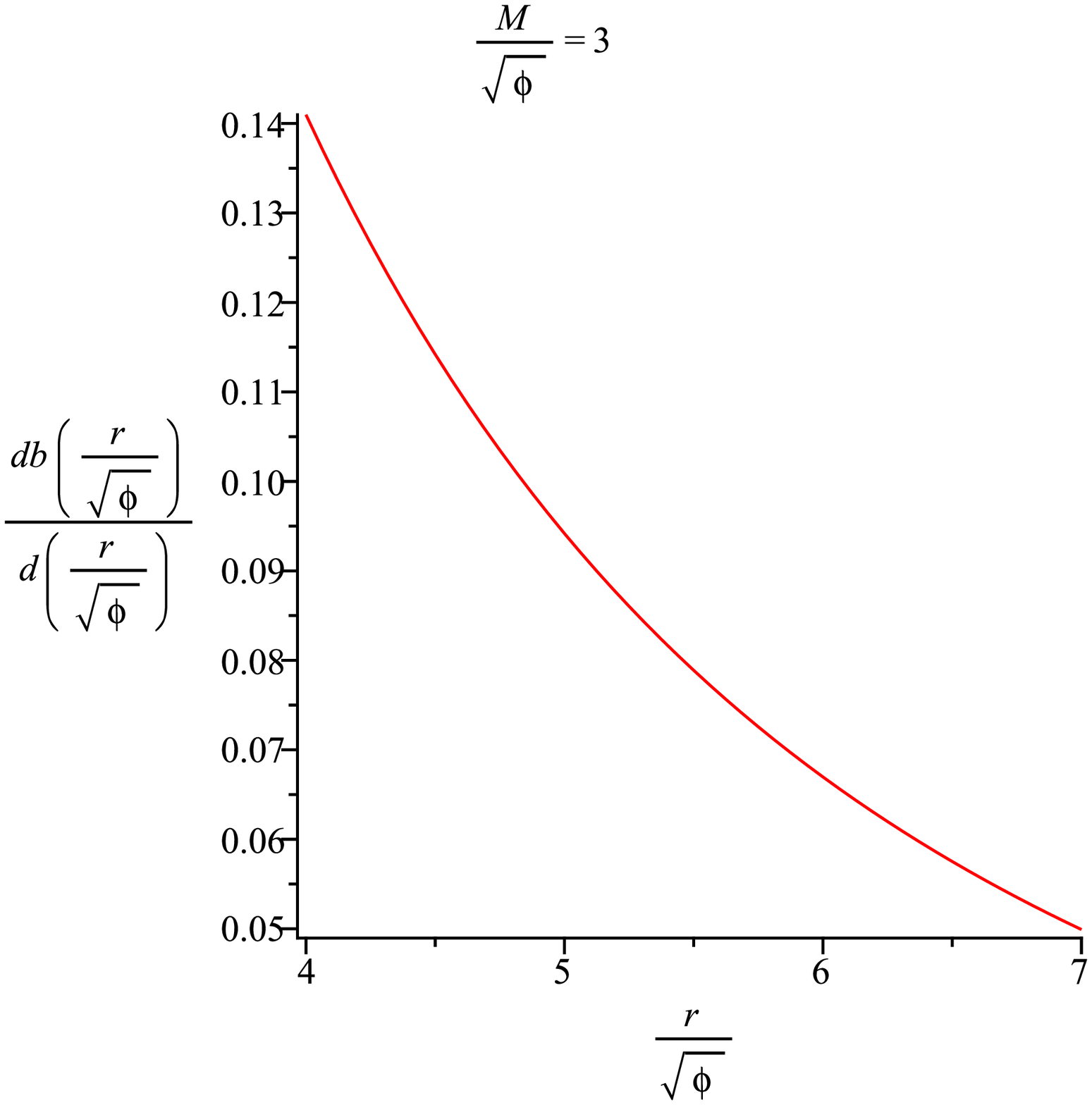}\\
 \\
\end{tabular}
\caption{(Left) Diagram of the shape function of the wormhole for
the specific value of the parameter $\frac{M}{\sqrt{\phi}}=3$.
(Middle) Diagram of the asymptotic behaviour of shape function.
(Right) Diagram of the derivative of the shape function of the
wormhole.    }
\end{figure*}

Now, we will discuss the behavioral effects of different aspects
of the above shape function $b(r)$ and its derivative. The throat
location of the wormhole is obtained by imposing the equation
$b(r_0) =r_0$. One can note that the appearance of a throat
depends on the parameter M and $\phi$. However, the larger root of
the equation $b\left(\frac{r_0}{\sqrt{\phi}}\right)
=\frac{r_0}{\sqrt{\phi}}$, where $\frac{r_0}{\sqrt{\phi}}$ is
dimensionless, gives the throat which depends only one parameter
$\frac{M}{\sqrt{\phi}}$. Figure 1 shows that the throat of  the
wormhole is located at   $\frac{r}{\sqrt{\phi}} =
\frac{r_0}{\sqrt{\phi}}$ (maximum root), where
$   \frac{r}{\sqrt{\phi}}-b\left(\frac{r_0}{\sqrt{\phi}}\right) $
cuts the   $\frac{r}{\sqrt{\phi}}$-axis. One can note that
position of the throat is increasing with the increase of smeared
mass distribution M. For $\frac{M}{\sqrt{\phi}}<2.2$ no throat
exists.  From the above analysis, we notice that we may get
feasible wormholes for $\frac{M}{\sqrt{\phi}}>2.2$.  For the sake
of brevity, we assume $\frac{M}{\sqrt{\phi}}=3$ for the rest of
the study.

From the left panel of Fig. 2, we observe that shape function is
increasing, therefore,
$b^{\prime}\left(\frac{r}{\sqrt{\phi}}\right)>0$. From Fig. 1,
 one can also note that for $\left(\frac{r}{\sqrt{\phi}}\right)
> \left(\frac{r_0}{\sqrt{\phi}}\right)$, $ \left(\frac{r}{\sqrt{\phi}}\right)-
b\left(\frac{r}{\sqrt{\phi}}\right)
 > 0$. This immediately implies that $\frac{b\left(\frac{r}{\sqrt{\phi}}\right)}{\left(
 \frac{r}{\sqrt{\phi}}\right)} < 1$
  which is an essential requirement for a shape function.
Right panel of figure 2 indicates that the flare-out condition
$b^{\prime}\left(\frac{r}{\sqrt{\phi}}\right) < 1$ for
$\left(\frac{r}{\sqrt{\phi}}\right)
> \left(\frac{r_0}{\sqrt{\phi}}\right)$ is satisfied.

We also observe the asymptotic behaviour from  the middle panel of
Fig. 2 such that
$\frac{b(\left(\frac{r}{\sqrt{\phi}}\right))}{\left(\frac{r}{\sqrt{\phi}}\right)}
\rightarrow 0$ as $\left(\frac{r}{\sqrt{\phi}}\right) \rightarrow
\infty$. Unfortunately, this has the similar explanation as done
in Ref. \cite{Rahaman2013b} that the redshift function does not
approach zero as $r \rightarrow \infty$ due to the conformal
symmetry. This means the wormhole spacetime is not asymptotically
flat and hence will have to be cut off at some radial distance
which smoothly joins to an exterior vacuum solution.

One can now find out the redshift function $f(r)$, where
$e^{2f(r)} = e^{\nu(r)}$. Using Eq. (7) we find
\begin {equation}
f(r)= ln(C_2 r).
\end{equation}

It can be observed from the above expression that the wormhole
presented here is traversable one as redshift function remains
finite.

The above solution should be matched with the exterior vacuum
spacetime of the Schwarzschild type at some junction interface
with radius $R$. Using this matching condition, one can easily
find the value of unknown constant $C_2$ as
\begin{equation}
C_2=\frac{e^{f(R)}}{R},
\end{equation}
so that the redshift function now explicitly becomes
\begin {equation}
f(r)= ln\left[\frac{r e^{f(R)}}{R}\right].
\end{equation}

The redshift function is therefore finite in the region $r_0 < r <
R$, as required because this will prevent an event horizon.
According to Fig. 3, $\phi(\rho+p_r)<0$, therefore, the null
energy condition is violated to hold a wormhole open.

\begin{figure}
      \centering  \includegraphics[scale=.45]{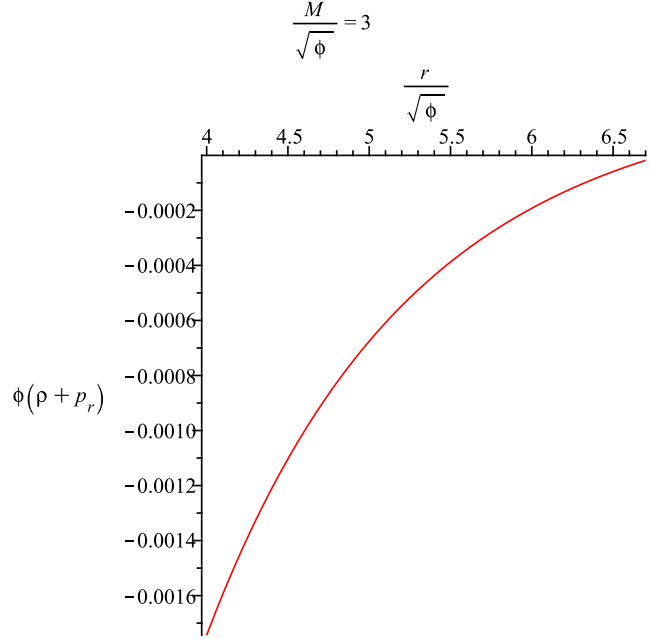}
        \caption{The violation of null  energy condition is shown against $\left(\frac{r}{\sqrt{\phi}}\right)$.}
   \label{fig:wh20}
\end{figure}

\subsection{THE TOLMAN-OPPENHEIMER-VOLKOFF EQUATION}

Following the suggestion of Ponce de Leon \cite{Leon1993}, we
write the Tolman-Oppenheimer-Volkoff (TOV) equation in the
following form
\begin{equation}
-\frac{M_G
(\rho+p_r)}{r^2}e^{\frac{\lambda-\nu}{2}}-\frac{dp_r}{dr}+\frac{2}{r}(p_t-p_r)=
0,
\end{equation}
where $M_G = M_G(r)$ is the effective gravitational mass within
the region from $r_0$ up to the radius $r$ and is given by
\begin{equation}
M_G(r)=\frac{1}{2}r^2
e^{\frac{\nu-\lambda}{2}}\nu^\prime.\label{mg}
\end{equation}

Equation (24) expresses the equilibrium condition for matter
distribution comprising the wormhole subject to the gravitational
force $F_g$, hydrostatic force $F_h$  plus another force $F_a$ due
to anisotropic pressure. Now, the above Eq. (24) can be easily
written as
\begin{equation}
F_g+ F_h+ F_a = 0,
\end{equation}
where
\begin{eqnarray}
F_g = -\frac{\nu^\prime}{2}\left(\rho +p_r\right)= -\frac{1}{4\pi
r^3} -\frac{M\sqrt{\phi}}{\pi^2 r(r^2+\phi)^2} \nonumber\\
-\frac{3M\sqrt{\phi}}{2\pi^2r^3(r^2+\phi)}+\frac{3M}{2\pi^2
r^4}tan^{-1}\left(\frac{r}{\sqrt{\phi}}\right),
\end{eqnarray}

\begin{eqnarray}
F_h = -\frac{dp_r}{dr}=\frac{1}{2\pi
r^3}+\frac{3M\sqrt{\phi}}{\pi^2 r^3(r^2+\phi)}
+\frac{3M\sqrt{\phi}}{\pi^2 r(r^2+\phi)^2}  \nonumber\\
-\frac{9M}{2\pi^2r^4}tan^{-1}\left(\frac{r}{\sqrt{\phi}}\right)+\frac{3M}{2\pi^2
r^3(r^2+\phi)},
\end{eqnarray}

\begin{eqnarray}
F_a = \frac{2}{r}(p_t-p_r)=-\frac{1}{4\pi
r^3}-\frac{2M\sqrt{\phi}}{\pi^2r(r^2+\phi)^2}  \nonumber\\
-\frac{3M\sqrt{\phi}}{\pi^2r^3(r^2+\phi)}+\frac{3M}{\pi^2
r^4}tan^{-1}\left(\frac{r}{\sqrt{\phi}}\right).
\end{eqnarray}

\begin{figure}
        \includegraphics[scale=.40]{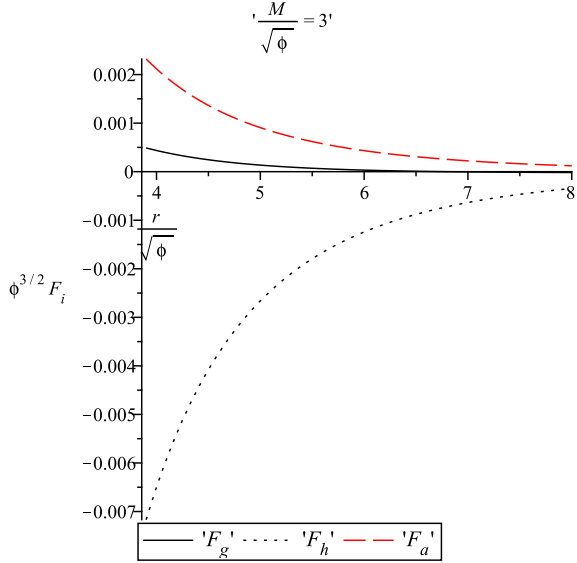}
        \caption{The variation of  $ \phi^{(3/2)}~\times$  forces are shown against $\frac{r}{\sqrt{\phi}}$.}
   \label{fig:F}
\end{figure}

From the Fig. \ref{fig:F} it can be observed that stability of the
system has been attained by gravitational and anisotropic forces
against hydrostatic force.

\subsection{ACTIVE GRAVITATIONAL MASS}

The active gravitational mass within the region from the throat
$r_0$ up to the radius $R$ can be found as
\begin{equation}
M_{active}= 4\pi \int^{  R  }_{  r_0   +} \rho r^2
dr=\frac{2M}{\pi}\left[tan^{-1}\left(\frac{r}{\sqrt{\phi}}\right)-
\frac{r\sqrt{\phi}}{r^2+\phi}\right]^ {  R  }_{
 r_0  +}.
\end{equation}

We observe here that the active gravitational mass $M_{active}$ of
the wormhole is positive under the constraint
$tan^{-1}\left(\frac{r}{\sqrt{\phi}}\right)
> \frac{r\sqrt{\phi}}{r^2+\phi}$ and also the nature of variation is
physically acceptable as can be seen from Fig. 5.

\begin{figure}
        \includegraphics[scale=.40]{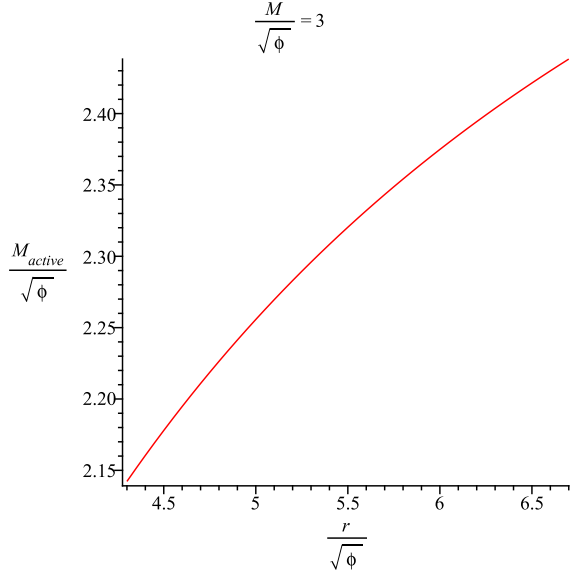}
        \caption{The variation of $\frac{M_{active}}{\sqrt{\phi}}$ is shown against
         $\frac{r}{\sqrt{\phi}}$.}
   \label{fig:wh20}
\end{figure}

\subsection{TOTAL GRAVITATIONAL ENERGY}

Using the prescription given by Lyndell-Bell et al.
\cite{Lyndell2007} and Nandi et al. \cite{Nandi2009}, we calculate
the total gravitational energy of the wormhole as
\begin{equation} E_g= Mc^2-E_M,
\end{equation}
where, $Mc^2= \frac{1}{2} \int_{r_0+}^R T_0^0 r^2 dr
+\frac{r_0}{2}$ is the total energy and $E_M=
\frac{1}{2}\int^R_{r_0+}  \sqrt{g_{rr}}  \rho r^2 dr$ is
 the total mechanical energy. Note that here $\frac{4 \pi}{8 \pi}$
  yields the factor $\frac{1}{2}$.

The range of the integration is considered here from the throat
 $\frac{r_0}{\sqrt{\phi}}$  to the embedded radial space of the wormhole geometry. We
have solved the above Eq. (31) numerically.

\begin{figure}
\centering
\includegraphics[scale=.40]{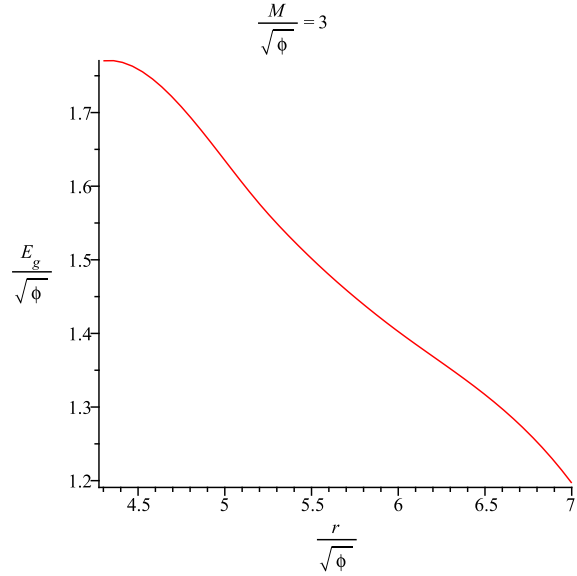}
\caption{The variation of $\frac{E_g}{\sqrt{\phi}}$ is shown
against $\frac{r}{\sqrt{\phi}}$.   Fig. 1 indicates that  for
$\frac{M}{\sqrt{\phi}} =3$,  $\frac{r_0}{\sqrt{\phi}} $ takes the value 4.275.}
\label{fig:E_g}
\end{figure}

\begin{table}
\caption{Data for plotting Fig. 6} \label{tab3} \centering
\bigskip
{\small

\begin{tabular}{@{}lcc@{}}
\hline \\[-9pt]
       Upper limit $R$             &Value of $E_g$\\

\hline \\[-9pt]
       5.0                         &1.635059734\\

\hline \\[-9pt]
       5.2                       &1.575910504\\

\hline \\[-9pt]
       5.4                        &1.524883013\\

\hline \\[-9pt]
       5.6                         &1.479809791\\

\hline \\[-9pt]
       5.8                        &1.439327656\\

\hline \\[-9pt]
       6.0                          &1.402514488\\

\hline \\[-9pt]
       6.2                         &1.368712897\\

\hline
\end{tabular}   }
\end{table}

In Fig. \ref{fig:E_g} we have considered
$\frac{M}{\sqrt{\phi}}=3$, throat radius
$\frac{r_0}{\sqrt{\phi}}=4.275$, the upper limit
 $\frac{R}{\sqrt{\phi}}$  is varying from 4.275+ to 7. 
 We have prepared a data sheet in Table 1 for plotting Fig. 6. 
 Here our observations are as follows: (1) When we are taking
$\frac{r_0}{\sqrt{\phi}}$ less than $4.275$, the value of the
integration become complex; (2) The numerical value of the
integration becomes real from the range of lower limit $4.275+$.
 These real and
positive values imply  $E_g > 0$, which  at once indicates that
there is a repulsion around the throat. Obviously this result is
expected for construction of a physically valid wormhole to
maintain stability of the fluid distribution.

\section{CONCLUDING REMARKS}

In the present letter we have considered anisotropic real matter
source for constructing new wormhole solutions. The background
geometry is inspired by noncommutativity along with conformal
Killing vectors to constrain the form of the metric tensor.
Speciality of this noncommutative geometry is to replace
point-like structure of gravitational source by smeared
distribution of the energy density under Gaussian distribution.
Notably, in this work we consider a point-like structure of
gravitational object without smearing effect where matter-energy
density is of the form provided by Mehdipour \cite{Mehdipour2012}.
Our investigation indicates that traversable wormhole solutions
exist in this Lorentzian distribution with physically interesting
properties under appropriate conditions.

The main observational highlights of the present study therefore
are as follows:

(1) The stability of the matter distribution comprising of the
wormhole has been attained in the present model. For this we have
calculated the TOV equation which expresses the equilibrium
condition for matter distribution subject to the gravitational
force $F_g$, hydrostatic force $F_h$ plus another force $F_a$ due
to anisotropic pressure.

(2) The active gravitational mass $M_{active}$ of the wormhole is
positive under the constraint
$tan^{-1}\left(\frac{r}{\sqrt{\phi}}\right)
> \frac{r\sqrt{\phi}}{r^2+\phi}$ as is expected from the
physical point of view.

(3) Since the total gravitational energy, $E_g > 0$, there is a
repulsion around the throat which is expected usually for stable
configuration of a wormhole.

As stated earlier in the text, in the present letter we employ
energy density given by Mehdipour \cite{Mehdipour2012} instead of
Nicolini-Smailagic-Spallucci type \cite{Nicolini2006}. However,
our overall observation is that in our present approach the
solutions and properties of the model are physically valid and
interesting as much as in the former approach. As a special
mention we would like to look at the Fig. 2 where we observe that
the shape function is increasing instead of monotone increase as
in the former case (Fig. 2 of Ref. \cite{Rahaman2013b}]. Further,
the redshift function does not approach zero as $r > r_0 $ due to
the conformal symmetry in both the approaches. So, exploration can
be done with some other rigorous studies between the two
approaches, i.e. Refs. \cite{Nicolini2006} and
\cite{Mehdipour2012}, which can be sought for in a future project.

\section*{Acknowledgments} 
F.R.and S.R.are thankful to the Inter-University Centre for 
Astronomy and Astrophysics (IUCAA), India for providing 
Visiting Research Associateship under which a part this work 
was carried out. IK is also thankful to IUCAA for research 
facilities. F.R.is grateful to UGC, India for financial support 
under its Research Award Scheme (Reference No.:  F.30-43/2011 (SA-II) ).
We are very grateful to an anonymous referee for his/her
insightful comments that have led to significant improvements,
particularly on the interpretational aspects.

\end{document}